\documentclass[twocolumn,nofootinbib,amsmath,amssymb,preprintnumbers]{revtex4}

\usepackage{graphicx}
\usepackage{slashbox}
\usepackage{color}
\usepackage{epstopdf}

\newcommand{\squishlist}{
   \begin{list}{$-$}
    { \setlength{\itemsep}{0pt}      \setlength{\parsep}{3pt}
      \setlength{\topsep}{4pt}       \setlength{\partopsep}{0pt}
      \setlength{\leftmargin}{1.5em} \setlength{\labelwidth}{1em}
      \setlength{\labelsep}{0.7em} } }

\newcommand{\squishlisttwo}{
   \begin{list}{-}
    { \setlength{\itemsep}{0pt}    \setlength{\parsep}{0pt}
      \setlength{\topsep}{0pt}     \setlength{\partopsep}{0pt}
      \setlength{\leftmargin}{2em} \setlength{\labelwidth}{1.5em}
      \setlength{\labelsep}{0.6em} } }

\newcommand{\squishend}{
    \end{list}  }

\newcommand{\tarttitle}[1]{``#1'',} 
\newcommand{\tbktitle}[1]{``#1''}     
\newcommand{\tISBN}[1]{#1} 

\newcommand{\bea}{\begin{eqnarray}}
\newcommand{\eea}{\end{eqnarray}}
\newcommand{\beq}{\begin{equation}}
\newcommand{\eeq}{\end{equation}}

\begin{document}
\title{Connecting the Micro-dynamics to the Emergent Macro-variables: Self-Organized Criticality and Absorbing Phase Transitions in the Deterministic Lattice Gas}
\author{Andrea Giometto$^{1,2}$}
\email{andrea.giometto10@imperial.ac.uk}
\author{Henrik Jeldtoft Jensen$^{3}$}
\email{h.jensen@imperial.ac.uk}
\affiliation{$^1$ 
Blackett Laboratory, Department of Physics and Complexity \& Networks Group, Imperial College London, London, SW7 2AZ, UK}
\affiliation{$^2$ Dipartimento di Fisica G. Galilei,  Universit\`a di Padova, Via Marzolo 8, I-35151 Padova, Italy}
\affiliation{$^3$ Department of Mathematics and Complexity \& Networks Group, Imperial College London, London, SW7 2AZ, UK}


\begin{abstract}
We reinvestigate the Deterministic Lattice Gas introduced  as a paradigmatic model of the $1/f$ spectra  (Phys. Rev. Lett. {\bf 64}, 3103 (1990)) arising according to the Self-Organized Criticality scenario. We demonstrate that the density fluctuations exhibit an unexpected dependence on systems size and relate the finding to effective Langevin equations. The low density behavior is controlled by the critical properties of the gas at the absorbing state phase transition. We also show that the Deterministic Lattice Gas is in the Manna universality class of absorbing state phase transitions. This is in contrast to expectations in the literature which suggested that the entirely deterministic nature of the dynamics would put the model in a different universality class. To our knowledge this is the first fully deterministic member of the Manna universality class. \\

\end{abstract}

\maketitle

\section{Introduction}
The aim of statistical mechanics is to establish a descriptive bridge between the micro world and the emergent observables at the macroscopic level. To comprehend the macroscopic world we make use of a few collective variables such as density or current. Ideally equations of motion for these variables should be rigorously derived from the equations of motion of the components, however this is seldom possible even for the simplest systems. An effective approach consists of semi-phenomenological Langevin equations\cite{vanKampen}. It is well known that although Langevin's approach is of great intuitive appeal it suffers from the fact that the details of the stochastic driving are very difficult to derive. Here we consider a microscopically {\em deterministic} model, namely the Deterministic Lattice Gas (DLG)\cite{HJJ_1} and show how a bulk noise term, absent for small system sizes\cite{HJJ_1,Grinstein1992}, is generated as the system size increases. In terms of driven diffusion equations it means that the behavior of the power spectrum of the density fluctuations can be understood in small systems as arising from a boundary driven diffusion equation\cite{HJJ_PS,Grinstein1992} without any Langevin bulk noise term. However as the system size is increased it turns out that a bulk noise term is needed in order to generate the correct power spectrum. The situation is particularly interesting because the model exhibits two regimes of very different behavior. At large density the fluctuations are well described by the appropriated diffusion equation. At low density the nearby absorbing state phase transition controls the fluctuations, which turns out to be in the Manna universality class.

The DLG is a model of a gas of particles undergoing overdamped dynamics. Numerical simulations and analytic studies concluded that the fluctuations in the number of particles exhibit robust $1/f$ fluctuations in several configurations. In the context of Self-Organized Criticality (SOC)\cite{Bak_1987}  this is of considerable interest since the original motivation for SOC was to establish a simple paradigm that might explain the widespread observation of $1/f$ temporal fluctuations and spatial fractals. DLG is thus an example of this scenario in contrast to the sandpile model used by Bak, Tang and Wiesenfeld to inaugurated which turned out to contain no $1/f$ spectra\cite{HJJ_book}.

\section{Model}
In the DLG particles interact through a nearest-neighbor repulsive central unit force and double occupancy  is not permitted.  All particles are updated simultaneously by moving the particles {\em deterministically} to neighbor sites   according to the vector sum of the forces they are subject to. If two particles want to move to the same site, the particle subject to the strongest force is moved, while in the case of equal forces no particle is displaced. Periodic boundary conditions are considered and we consider the model in dimension $d=2$.

The number of particles $N(t)$ in a sub-volume of the lattice exhibits interesting temporal fluctuations.
We determine the power spectrum $S(f)$ of $N(t)$ from the square of the absolute value of the Fourier transform. Successive time sequences are averaged in order to achieve sufficient statistics:
\beq
S(f) = \frac{1}{2 \pi T} \left\langle \left\vert \sum_{t=1}^{T} N(t) e^{-i 2 \pi f t} \right\vert^2 \right\rangle
\label{S(f)_measured}
\eeq
The angular brackets denote averaging over many different time series.

\begin{figure}[!hbt]
\begin{center}
\includegraphics[width=6cm,angle=-90]{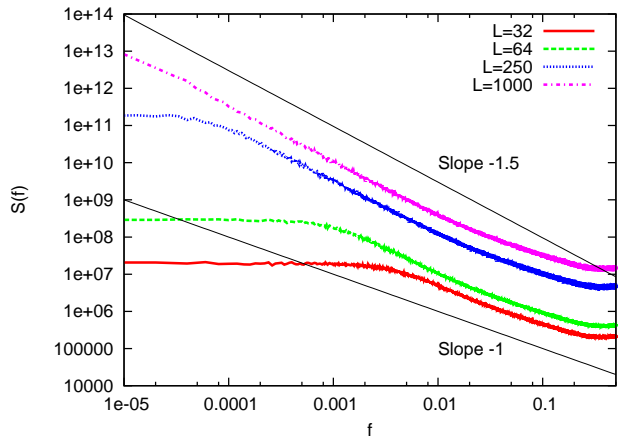}
\caption{Scaling behavior of the spectrum $S(f) \propto f^{-\mu}$ of the total number of particles $N(t)$ in the DLG for increasing linear sizes of the lattice $L$. A crossover from $\mu = 1$ for small $L$ to $\mu = 1.5$ for large $L$ is observed. Particle density $\rho=0.5$.}
\label{S_size_depen}
\end{center}
\end{figure}

The spectrum $S(f)$ was shown\cite{HJJ_diff} to satisfy $S(f) \sim 1/f^\mu$ at the particle density $\rho = 0.3$, while previous works showed that the same result is obtained in a wide range of densities with a drive at the boundary\cite{HJJ_1}. In those papers the maximum linear system size considered is $L=128$, we will see that different behaviors are observed for larger systems. In Fig. \ref{S_size_depen} we show how the effective exponent $\mu$ changes from $\mu=1$ to $\mu=3/2$ as $L$ increases. It is interesting to relate this change in the temporal fluctuations to the appropriate macroscopic equation of motion for the particle density $\rho({\bf r},t)$.

Simulations show that although the particles move deterministically in the DLG model they do behave like random walkers in the sense that their square displacement is linear in time. This suggests that $\rho({\bf r},t)$ evolves according to a diffusion equation and integrating the diffusion equation allows us to extract the temporal evolution of $N(t)=\int_V d{\bf r} \rho({\bf r},t)$, where $V$ denotes the measuring sub-volume. Assume that $\rho({\bf r},t)$ satisfies the diffusion
\beq
\frac{\partial \rho}{\partial t} = D\nabla^2\rho + \xi.
\label{diffuse}
\eeq
It has been shown\cite{Grinstein1992,HJJ_book} that the power spectrum of $N(t)$ depends on the details of the diffusion equation. If the equation is driven by white noise on the boundary of $V$ and the noise term $\xi$ is absent the power spectrum exponent derived from Eq. (\ref{diffuse}) is $\mu=1$. If in contrast a conservative bulk noise term is included in Eq. (\ref{diffuse}) one obtains instead $\mu=3/2$. Including non-linearities in Eq. (\ref{diffuse}) will not influence the behavior of $\mu$, see \cite{Grinstein1992}. It is this observation that allows us to conclude that a conservative noise term is generated in the macroscopic Langevin description when the sub-volume $V$ becomes sufficiently large. 

\section{$S(f)$ scaling for large lattices}
\begin{figure}[!hbt]
\begin{center}
\includegraphics[width=6cm,angle=-90]{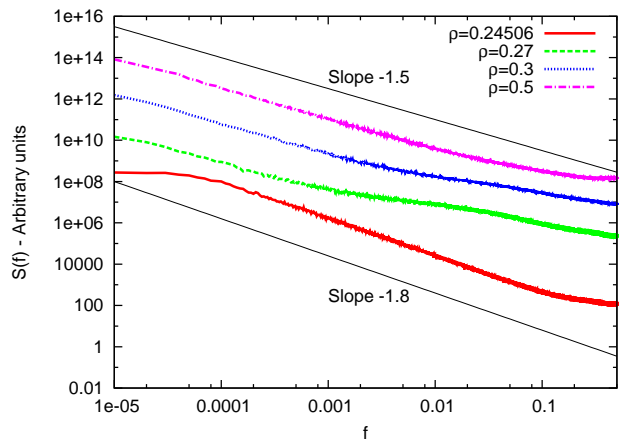}
\caption{Scaling behavior of the spectrum $S(f) \propto f^{-\mu}$ of the total number of particles $N(t)$ in the DLG for different particle densities $\rho$. A crossover from $\mu \simeq 1.8$ at $\rho \simeq \rho_c=0.245$ to $\mu = 1.5$ at $\rho \gg \rho_c$ is observed. $S(f)$ has been multiplied by different constants for different densities $\rho$ to visualize the scaling exponents properly. Lattice linear size $L=1000$.}
\label{spectra_pdlg}
\end{center}
\end{figure}

We now consider the DLG for much larger lattices than previously studied and our aim is to understand the fluctuations spectrum as a function of density.  We plot $S(f)$ in Figure \ref{spectra_pdlg} and find that the value of $\mu$  depends on the density:
\begin{equation}
 S(f) \sim
 \begin{cases}
 f^{-1.5} & \rho \geq 0.3 \\
 f^{-1.8} & \rho \simeq \rho_c = 0.245
 \end{cases}
 \end{equation}
where $\rho_c$ is the critical density of the absorbing phase transition (APT in the following).
The observed exponent at high densities is the same of a gas of random walkers\cite{HJJ_2,HJJ_3}, while  at low densities the power spectrum scaling exponent is determined by the critical properties of the DLG at the APT, as we shall discuss in the following.

\section{Absorbing Phase Transition}
At very low densities the DLG enters configurations in which all particles are far away from each other and, due to the short range interactions, the particles become unable to move, i.e. the dynamics is frozen. This is called an absorbing state phase transition between an active phase, above the critical density $\rho_c$, and an inactive below.

The control parameter of the transition is the particle density $\rho$, while the order parameter is the density of active particles $\rho_a$ (this is the average fraction of particles that move in one time step). Near the transition:
\begin{equation}
 \rho_a(\delta\rho) \sim
 \begin{cases}
 \delta\rho^{\beta} & \rho > \rho_c \\
 0 & \rho < \rho_c
 \end{cases}
 \end{equation}
 with $\delta\rho=\rho-\rho_c$. The determination of the critical density $\rho_c$ is obtained by varying its value until one observes a straight line in a log-log plot, which gives $\rho_c=0.24500(2)$. The brackets indicate the statistical uncertainty on the last digit, which is estimated by looking at deviations from a straight line in the same plot. A regression analysis yields the value of the order parameter exponent $\beta=0.634(2)$.
\begin{figure}[!hbt]
\begin{center}
\includegraphics[width=6cm,angle=-90]{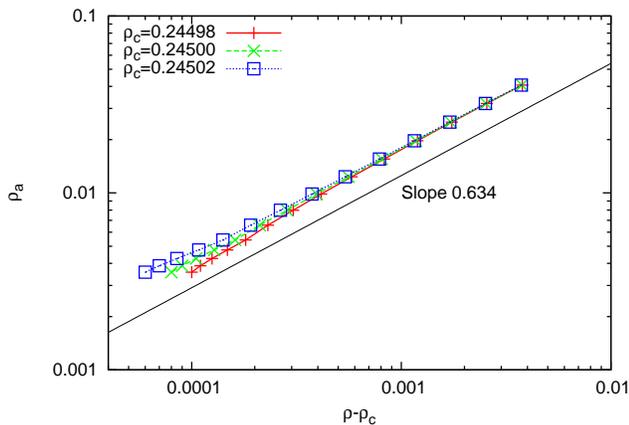}
\caption{Density of active sites $\rho_a$ as a function of $\delta \rho = \rho - \rho_c$. The determination of the critical density $\rho_c$ is obtained varying its value until data points are aligned on a straight line in a log-log plot. Lattice linear size $L=1000$. Error bars are smaller than symbols.}
\label{rhoc}
\end{center}
\end{figure}

\begin{figure}[!hbt]
\begin{center}
\includegraphics[width=6cm,angle=-90]{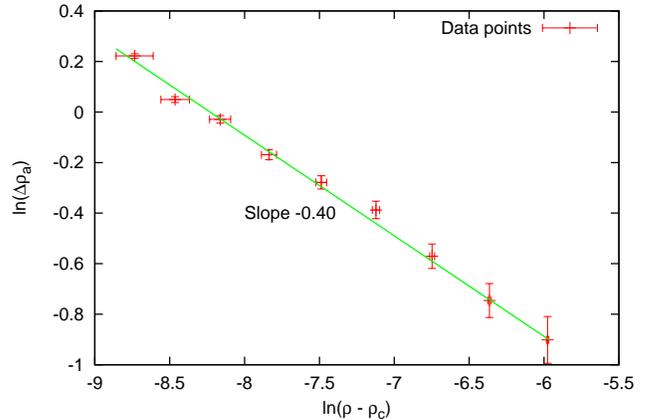}
\caption{Determination of the exponent $\gamma'$. Lattice linear size $L=1000$.}
\label{gammafirst}
\end{center}
\end{figure}

The fluctuations of the order parameter $\Delta \rho_a = L^2 \left[ \langle \rho_a^2 \rangle - \langle \rho_a \rangle^2 \right]$ are plotted in Fig. \ref{gammafirst}. Approaching the transition point $\Delta \rho_a$ increases and diverges at $\rho_c$. Close to the critical point fluctuations scale as
\begin{equation}
\Delta\rho_a \sim \delta\rho^{-\gamma'}
\label{scaling_fluct}
\end{equation}
in the active phase. Using a regression analysis one gets $\gamma'=0.40(2)$.

\subsection{External field}
\begin{figure}[!hbt]
\begin{center}
\includegraphics[width=6cm,angle=-90]{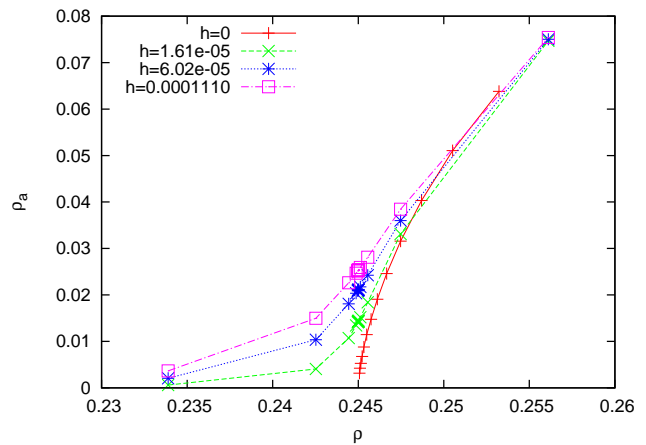}
\caption{Density of active sites $\rho_a$ vs particle density $\rho$ for different values of the external field $h$. At $h>0$ the system does not fall into an absorbing state. Error bars are smaller than symbols. Lattice linear size L=1000.}
\label{heffect}
\end{center}
\end{figure}

A feature of finite size effects is that the system may pass within one simulation from one phase to the other. This is caused by critical fluctuations, that diverge while approaching the critical point: as soon as the correlation length $\xi_{\perp}$ is of the order of $L$ the system may pass to the absorbing phase even in a small region above the critical point $\rho_c$.
Similar to equilibrium phase transitions it is possible to apply an external field $h$ conjugated to the order parameter, i.e., in our case the field corresponds to spontaneous creation of active particles, which prevents the system to enter an absorbing state and thus allows to study scaling properties at the critical point $\rho=\rho_c$. Here we make use of the implementation of the external field for APTs that was developed in \cite{Lubeck_2}.  Since the control parameter $\rho$ is a conserved quantity, the external field must preserve the conservation: at each time step, after performing the DLG update, we choose randomly $h L^2$ particles on the lattice and move each of them to one of its empty neighbors. In this way inactive particles may be activated and the number of active sites is increased, while the density of particles $\rho$ is conserved. The introduction of the external field prevents the system from falling into an absorbing state and the order parameter $\rho_a$ does not vanish at $\rho = \rho_c$, as can be seen in Fig. \ref{heffect}.

The order parameter and its fluctuations are assumed to be generalized homogeneous functions of the control parameter $\rho$ and of the applied field $h$:
\beq
\rho_a (\delta \rho, h) = \lambda^{-\beta} \tilde R (\delta\rho \lambda, h \lambda^{\sigma})
\label{ansatz_rhoa}
\eeq
\beq
\Delta \rho_a (\delta \rho, h) = \lambda^{\gamma'} \tilde D (\delta\rho \lambda, h \lambda^{\sigma})
\label{ansatz_fluct}
\eeq
where $\lambda>0$. From equation (\ref{ansatz_rhoa}) one can obtain an estimate of the exponent $\sigma$ plotting $\rho_a$ against $h$ in a log-log plot:
\begin{equation}
\rho_a(\delta\rho=0,h)\sim h^{\beta/\sigma}
\end{equation}
This gives $\sigma=2.19(1)$\footnote{The error on $\beta/\sigma$ is computed as the dispersion on its value when measured at $\rho_{c,1}=0.24498$ and $\rho_{c,2}=0.24502$ (respectively the lower and upper bound on our determination of the critical density $\rho_c=0.24500(2)$). Errors on other quantities are estimated in the same way, when possible.}.
%
%
An independent estimate of $\gamma'$ can be obtained through equation (\ref{ansatz_fluct}) in the same way:
\begin{equation}
\Delta\rho_a(\delta\rho=0,h)\sim h^{-\gamma'/\sigma}
\end{equation}
which gives $\gamma'=0.365(7)$. We consider the weighted average of the two determinations of $\gamma'$ as our best estimate: $\gamma'=0.37(1)$.

\begin{figure}[!hbt]
\begin{center}
\includegraphics[width=6cm,angle=-90]{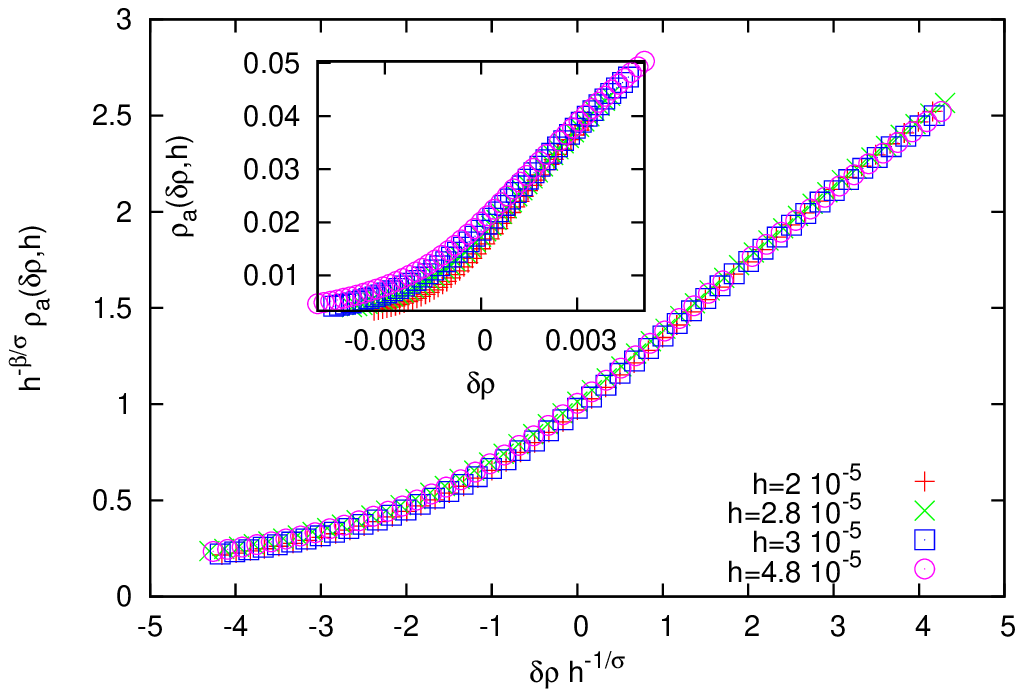}
\caption{Data collapse as in equation (\ref{collapse_rhoa_eq}). In the inset data before the collapse are shown. Lattice linear size $L=1000$.}
\label{collapse_rhoa}
\end{center}
\end{figure}

\begin{figure}[!hbt]
\begin{center}
\includegraphics[width=6cm,angle=-90]{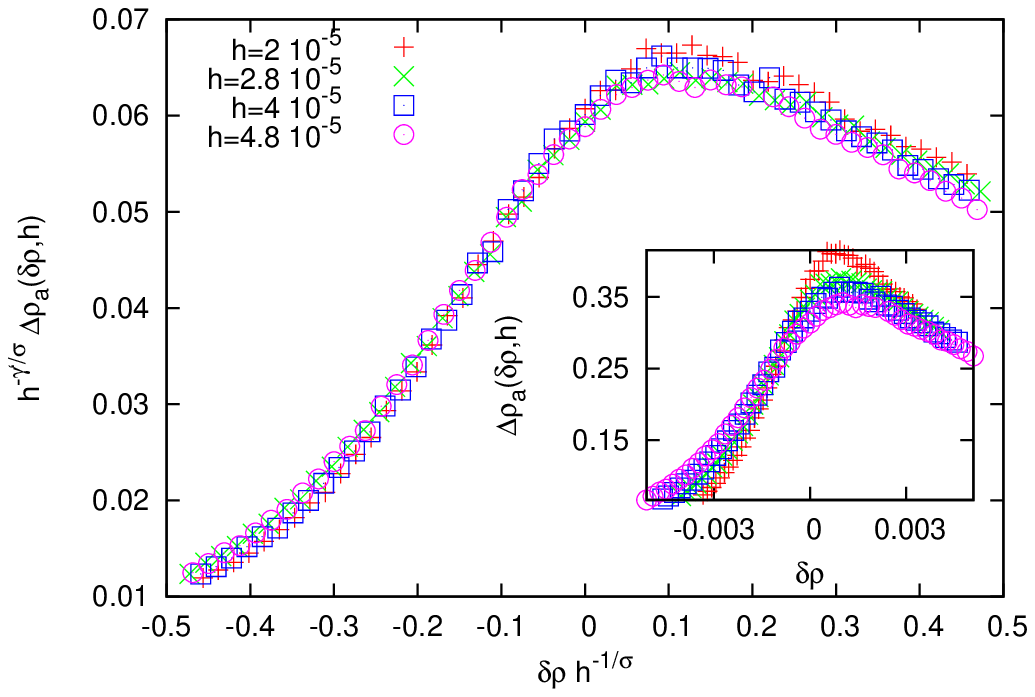}
\caption{Data collapse as in equation (\ref{collapse_fluct_eq}). In the inset data before the collapse are shown. Lattice linear size $L=1000$.}
\label{collapse_fluct}
\end{center}
\end{figure}

Data collapses can be produced choosing $\lambda=h^{-1/\sigma}$ in equations (\ref{ansatz_rhoa}) and (\ref{ansatz_fluct}), see figures \ref{collapse_rhoa} and \ref{collapse_fluct}:
\begin{equation}
h^{-\beta/\sigma} \rho_a (\delta \rho,h) = \tilde R(\delta \rho h^{-1/\sigma}, 1)
\label{collapse_rhoa_eq}
\end{equation}
\begin{equation}
h^{\gamma'/\sigma} \Delta \rho_a (\delta\rho, h) = \tilde D(\delta \rho h^{-1/\sigma}, 1)
\label{collapse_fluct_eq}
\end{equation}

\subsection{Finite size scaling}
Similar to equilibrium critical phenomena we assume that the system size L enters the scaling forms (\ref{ansatz_rhoa}) and (\ref{ansatz_fluct}) as an additional scaling field, i.e.,
\beq
\rho_a (\delta \rho, h, L) = \lambda^{-\beta} \tilde R_{pbc} (\delta\rho \lambda, h \lambda^{\sigma}, L \lambda^{-\nu_{\perp}})
\label{ansatz_rhoa_finite}
\eeq
\beq
\Delta \rho_a (\delta \rho, h, L) = \lambda^{\gamma'} \tilde D_{pbc} (\delta\rho \lambda, h \lambda^{\sigma}, L \lambda^{-\nu_{\perp}})
\label{ansatz_fluct_finite}
\eeq
where the exponent $\nu_{\perp}$ describes the divergence of the spatial correlation length, i.e., $\xi_{\perp} \propto \delta \rho^{-\nu_{\perp}}$. The universal scaling functions depend on the particular choice of the boundary conditions, although in the thermodynamic limit $\tilde R_{pbc}(x,y,\infty)=\tilde R(x,y)$ and $\tilde D_{pbc}(x,y,\infty)=\tilde D(x,y)$.

\begin{figure}[!hbt]
\begin{center}
\includegraphics[width=6cm,angle=-90]{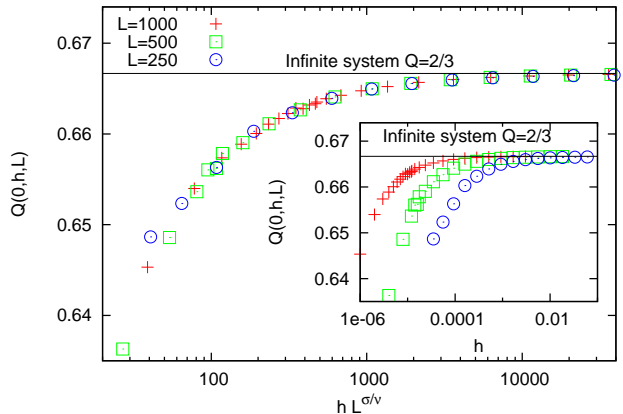}
\caption{(lin-log) Data collapse as in equation (\ref{nu}) for the determination of $\nu_{\perp}$. In the inset data before the collapse are shown.}
\label{nu_collapse}
\end{center}
\end{figure}

Following \cite{Lubeck_5} we consider the fourth order cumulant $Q$, which is defined as:
\beq
Q = 1 - \frac{\langle \rho_a^4 \rangle}{3 \langle \rho_a^2 \rangle^2}
\label{Q}
\eeq
For non-vanishing order-parameter the cumulant tends to $Q=2/3$ in the thermodynamic limit. One expects it to obey the scaling form:
\beq
Q(\delta \rho, h, L) = \tilde Q_{pbc} (\delta \rho \lambda, h \lambda^{\sigma}, L \lambda^{-\nu_{\perp}})
\label{Q_scaling}
\eeq
Choosing $\lambda=L^{1/{\nu_{\perp}}}$ in equation (\ref{Q_scaling}) at $\delta \rho = 0$ we obtain the following equation:
\begin{equation}
Q(0, h, L) = \tilde Q_{pbc} (0, h L^{\sigma/{\nu_{\perp}}}, 1)
\label{nu}
\end{equation}
which enables us to determine $\nu_{\perp}$ through a data collapse by plotting $Q(0,h,l)$ against ${hL^{\sigma/\nu_{\perp}}}$ as in Fig. \ref{nu_collapse}. Best results are obtained for $\nu_{\perp}=0.83(5)$\footnote{The error is estimated looking at data collapses for different values of the exponent.}.

\subsection{Dynamical scaling}
Starting simulations of the DLG from a random distribution of particles above the critical point $\rho_c$ the density of active sites $\rho_a$ decreases in time and tends to its steady state value. At the critical point $\rho=\rho_c$ the order parameter decays algebraically as:
\begin{equation}
\rho_a(\delta \rho=0, h=0,t) \sim t^{-\alpha}
\label{alpha}
\end{equation}
Simulating a lattice of linear size $L=4000$ at $\rho = \rho_c$ we obtain $\alpha=0.41(1)$.

\begin{figure}[!hbt]
\begin{center}
\includegraphics[width=6cm,angle=-90]{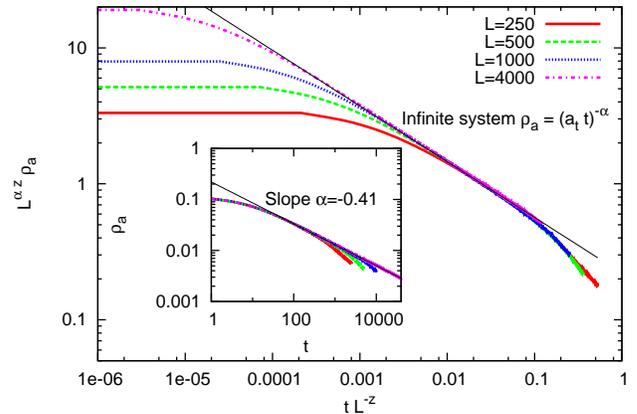}
\caption{Data collapse as in equation (\ref{zeta_2}) to estimate $z$. In the inset data before the collapse are plotted.}
\label{collapse_z}
\end{center}
\end{figure}

As can be seen in the inset of Fig. \ref{collapse_z}, a finite system size limits this power law behavior and one expects the following ansatz to hold at criticality:
\beq
\rho_a(L,t) = \lambda^{-\alpha \nu_{\parallel}} \tilde R_{pbc}' (t \lambda^{- \nu_{\parallel}}, L \lambda^{-\nu_{\perp}})
\label{zeta}
\eeq
Setting $a_L L \lambda^{-\lambda_{\nu_{\perp}}}=1$ one obtains:
\beq
\rho_a(L,t) = L^{-\alpha z} \tilde R_{pbc}'(t L^{-z},1)
\label{zeta_2}
\eeq
with $z = \nu_{\parallel}/\nu_{\perp}$. Through equation (\ref{zeta_2}) one can obtain an estimate of $z$ performing a data collapse, plotting $L^{\alpha z}\rho_a$ against $tL^{-z}$ as in Fig. \ref{collapse_z} and varying the value of the exponent until the collapse is satisfactory. The best data collapse is obtained for $z=1.5(1)$\footnote{The error is estimated looking at data collapses for different values of the exponent.}.

\subsection{DLG universality class}
After having determined the scaling exponents we are able to address the question concerning which universality class the DLG belongs to. In table \ref{DLG_Manna} we compare the measured values of the DLG critical exponents with those of the Manna universality class\cite{Lubeck_book}, showing that they are compatible with each other\footnote{The estimate of $\nu_\parallel$ is obtained from $z=\nu_\parallel/\nu_\perp$, knowing $z$ and $\nu_\perp$. The value of $\gamma$ is obtained through Widom law ($\gamma = \sigma - \beta$).}. Some of the errors in our determinations of the exponent are significantly larger than the errors on the estimates of the Manna universality class available in the literature: this is probably due to the fact that we cannot afford to perform as many iterations of the model (to compute ensemble average) as those performed for the CLG, CTTP and Manna model\cite{Lubeck_book}, since the DLG is computationally very demanding to simulate. It is also likely that stochastic models are more self-averaging than our deterministic model is and therefore allows more accurate estimates of the exponents in less iterations.
\begin{table}[!hbt]
\begin{center}
\begin{tabular}{|c|c|c|c|c|}	\hline
	&	$\beta$		&	$\nu_\perp$	&	$\nu_\parallel$	&	$\sigma$\\	\hline
Manna	&	0.639(9)	&	0.799(14)	&	1.225(29)	&	2.229(32)	\\
DLG		&	0.634(2)	&	0.83(5)	&	1.2(1)	&	2.19(1)\\	\hline	\hline
	&	$\gamma'$	&	$\gamma$	&	$\alpha$	&	$z$	\\	\hline
Manna	&	0.367(19)	&	1.590(33)		&	0.419(15)		&	1.533(24)	\\
DLG 		&	0.37(1)	&	1.54(1)		&	0.41(1)		&	1.5(1)	\\ \hline
\end{tabular}
\caption{The measured critical exponents for the DLG and the corresponding critical exponents for the Manna universality class in $d=2$\cite{Lubeck_book}.}
\label{DLG_Manna}
\end{center}
\end{table}
The shape of the universal scaling functions is remarkably similar to those that can be found in the literature for the Manna universality class. This is seen e.g. by comparing our Fig. \ref{collapse_fluct} with Fig. 4 in \cite{Lubeck_2}, see also \cite{Lubeck_book}. From the scaling exponents and the scaling functions we find compelling evidence that the DLG belongs to the Manna universality class. 
 
\section{Power spectrum at criticality}
In this section we want to relate the exponent of the power spectrum at low density to the scaling properties of the DLG at the critical point $\rho_c$ of the APT\cite{Lauritsen}. As we will see the scaling behavior of the correlation function $C_a(\mathbf{r},t)=\langle \rho_a(\mathbf{r},t) \rho_a(0,0) \rangle - \langle \rho_a(0,0)^2 \rangle$ at the critical density $\rho=\rho_c$ determines the power law exponent in the scaling of the spectrum $S_a(f)$ of the total number of \textit{active} particles $N_a(t)$ in the lattice.

\begin{equation}
C_a(\mathbf{r},t) = \lambda^{-\eta} \tilde C_a (\lambda \rho, \lambda^{-\nu_\parallel} t, \lambda^{-\nu_{\perp}} r)
\end{equation}
 
In stationary directed percolation processes above criticality it has been observed\cite{Lubeck_book} that the correlation function $C_a(\mathbf{r},t)$ at $t=0$ first decays in space algebraically as $r^{-\beta/\nu_\perp}$, until it saturates at a constant value at $r > \xi_\perp$. In the saturated regime the two sites become uncorrelated so that this value is just equal to the squared stationary density of active sites $\rho_a^2$.

One can use this fact to extract the scaling exponent for $S_a(f)$, making use of the Wiener-Khinchin theorem:
\begin{align*}
C_a(t) &= \langle N_a(t) N_a(0) \rangle - \langle N_a(0)^2 \rangle \\ 
&= \left\langle \int_V d^2\mathbf{r} \int_V d^2\mathbf{r'} \rho_a(\mathbf{r},t) \rho_a(\mathbf{r'},0) \right\rangle + \dots \\
&= \left\langle V \int_V d^2\mathbf{r} \ \rho_a(\mathbf{r},t)\rho_a(0,0) \right\rangle + \dots \ \\ &= \ \int_V d^2\mathbf{r} \ C_a(\mathbf{r},t) + \dots 
\sim \int_V dr \ r \ r^{-\beta/\nu_\perp} \tilde C_a(t/r^z) \ \\
& \sim \ t^{1/z(2-\beta/\nu_\perp)} 
\end{align*}
so that:
\beq
S_a(f) = \frac1{2\pi }\int dt \ C_a(t) \ e^{-i 2 \pi f t} \sim f^{-1-\frac{1}{z}\left(2-\frac{\beta}{\nu_\perp}\right)} \sim f^{-\mu}
\eeq
with $\mu = 1+\frac{1}{z}\left(2-\frac{\beta}{\nu_\perp}\right)$. With our estimates of the critical exponents for the DLG we find $\mu=1.82(6)$, while with the best estimates of the Manna universality class exponents\cite{Lubeck_book} one finds $\mu=1.78(2)$.

\begin{figure}[!hbt]
\begin{center}
\includegraphics[width=6cm,angle=-90]{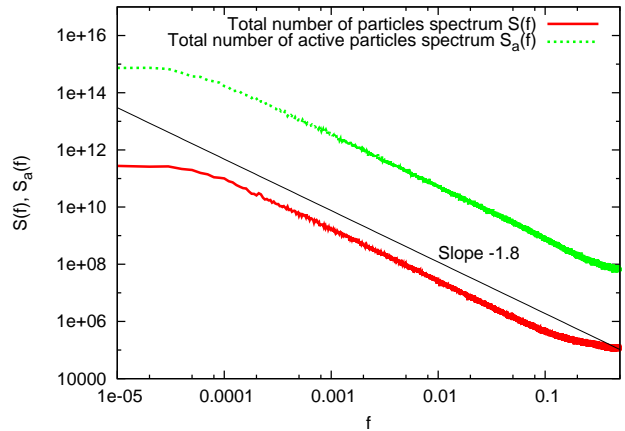}
\caption{DLG: Scaling behavior of the spectrum $S(f) \propto f^{-\mu}$ of the total number of particles in the box $N(t)$ and of the spectrum $S_a(f)$ of the total number of active particles near the critical density $\rho=0.24506 \simeq \rho_c$. The spectra have been multiplied by arbitrary factors to visualize the scaling exponent properly. Lattice linear size $L=1000$.}
\label{spectra_pdlg_critical}
\end{center}
\end{figure}

At low densities we expect that fluctuations in the total number of particles $N(t)$ in the lattice are triggered by active particles, therefore we expect the two spectra $S(f)=\frac1{2\pi}\int dt \ N(t) \ e^{-i 2 \pi f t}$ and $S_a(f) =\frac1{2\pi}\int dt \ N_a(t) \ e^{-i 2 \pi f t}$ to show the same scaling behavior at $\rho \simeq \rho_c$, as it is confirmed by simulations (see Fig. \ref{spectra_pdlg_critical}).

Computing the power spectrum of the total number of particles in the box at $\rho=0.24506 \simeq \rho_c$ we observe a power spectrum compatible with the predictions, confirming that fluctuations in the total number of particles in the sub-volume $V$ are determined by the scaling behavior of the correlation function at criticality.

\section{Conclusions}
We have investigated the scaling properties of the power spectrum $S(f)$ of the total number of particles in a sub-volume of the lattice for the DLG: our simulations show that the DLG is not characterized by $1/f$ fluctuations as had been previously observed, but present a much more variegated picture. At high densities the power spectrum is the same as observed in a gas of random walkers, while at low densities the spectrum scales as $S(f) \sim 1/f^{1.8}$ and we have shown that this exponent is determined by the decay of the density-density  correlation function near criticality.

Furthermore we have shown that the deterministic lattice gas is in the same universality class as the Manna model\cite{Manna}. To our knowledge this is the first example of a completely deterministic and non-chaotic system in this universality class. Perhaps this is not too surprising given that we found that a stochastic bulk noise is generated in the effective Langevin description as the system size is increased. From this point of view the DLG becomes similar to the other members of the Manna class, which all involved a stochastic element in their microscopic dynamics. Examples of models in the Manna class includes the Oslo model\cite{Oslo_Model} and the Manna model\cite{Manna}. In the Oslo model the local threshold for relaxation is updated stochastically at every relaxation and in the Manna model particles move to stochastically chosen neighbor sites. 

 Further studies could focus on the critical behavior of the DLG in one, three and higher dimensions, or on deterministic variants of the Manna and Conserved Threshold Transfer models\cite{Rossi}, to address the question of how a deterministic micro-dynamics affects the critical behavior of these systems. Recently a method\cite{Bonachela} has been devised to check wether a model belongs to the Manna or to the DP universality class, that have similar critical exponents. This method, that involves the study of critical properties near an absorbing or reflecting wall, could be used to confirm the fact that the DLG belongs to the Manna class.

\section*{Appendix}
\section*{Non-universal metric factors} When we express the order parameter and other quantities as generalized homogeneous functions like in equations (\ref{ansatz_rhoa}) or (\ref{ansatz_fluct}) we can introduce the non-universal metric factors $a_\rho$, $a_\Delta$ and $a_h$ so that once they are chosen in a specified way, the universal scaling functions are the same for all systems within the same universality class, while these factors are non-universal in the sense that they can be different for models in the same class (see \cite{Lubeck_book} for further details):
\beq
\rho_a (\delta \rho, h) = \lambda^{-\beta} \tilde R (a_{\rho} \delta\rho \lambda, a_h h \lambda^{\sigma})
\eeq
\beq
a_\Delta \Delta \rho_a (\delta \rho, h) = \lambda^{\gamma'} \tilde D (a_{\rho} \delta\rho \lambda, a_h h \lambda^{\sigma})
\eeq
In table \ref{non-univ_DLG} we report the measured values of the non-universal metric factors for the DLG.
\begin{table}[!hbt]
\begin{center}
\begin{tabular}{|c|c|c|c|c|}	\hline
	&	$a_\rho$		&	$a_\Delta$	&	$a_h$	&	$a_t$		\\	\hline
DLG	&	1.741(9)	&	27(2)	&	0.028(2)	&	40.1(1)	\\
CLG	&	0.509	&	9.241	&	0.062	&	11.22	\\
CTTP	&	0.341	&	45.42	&	0.013	&	24.90	\\
Manna	&	0.211	&	78.56	&	0.007	&	35.53	\\ \hline
\end{tabular}
\caption{The measured non-universal metric factors for the DLG and for the CLG, CTTP and Manna models\cite{Lubeck_book} (the uncertainty on these values is less than $5\%$), in $2d$.}
\label{non-univ_DLG}
\end{center}
\end{table}

\section*{Tracer diffusion constant as an order parameter}
It has been recently observed\cite{Diffusion_sandpiles} that in some one-dimensional stochastic sand-piles the tracer diffusion coefficient $D_t$ ($\langle R^2(t) \rangle = 2 \ D_t t$, where $R(t)$ is the displacement of a particle at time $t$) scales in the same manner as the density of active particles, so that it represents an alternative definition of an order parameter. Here we show that the same behavior is found in the DLG, which enables us to obtain independent estimates of the scaling exponents $\beta$ and $\sigma$.

\begin{figure}[!htb]
\begin{center}
\includegraphics[width=6cm,angle=-90]{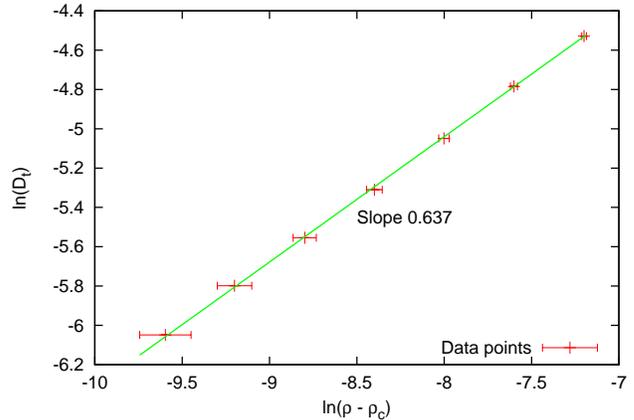}
\caption{Determination of the order parameter exponent $\beta$ through the scaling of the tracer diffusion coefficient: $D_t(\delta\rho,h=0) \sim \delta\rho^\beta$. $L=1000$.}
\label{beta_Dt}
\end{center}
\end{figure}

Scale invariance implies that the new order parameter $D_t$ can be written as a generalized homogeneous function (for an infinite system):
\begin{equation}
D_t(\delta\rho,h)=\lambda^{-\beta} \tilde T (\delta\rho \lambda, h \lambda^\sigma)
\end{equation}
so that we expect the following ansatz to be satisfied at $h=0$ near the critical point $\rho_c$:
\begin{equation}
D_t(\delta\rho,h=0) \sim \delta\rho^\beta
\end{equation}
This is verified in Fig. \ref{beta_Dt} and a linear regression analysis gives $\beta=0.637(4)$.

\begin{figure}[!htb]
\begin{center}
\includegraphics[width=6cm,angle=-90]{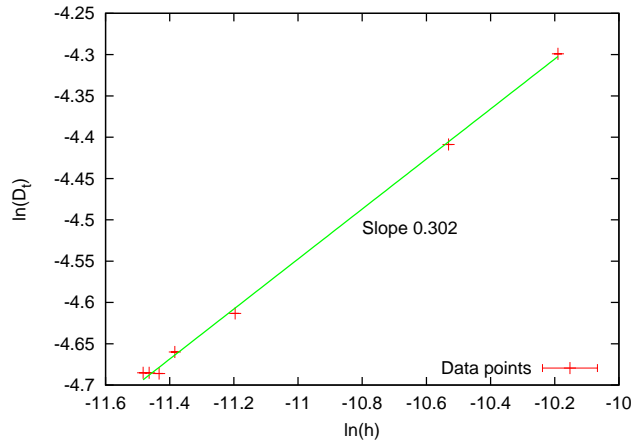}
\caption{Determination of $\beta/\sigma$ through the scaling of the tracer diffusion coefficient in the presence of an external field $h$: $D_t(\delta\rho=0,h) \sim h^{\beta/\sigma}$. Error bars are smaller than symbols. $L=1000$.}
\label{Dt_vs_h}
\end{center}
\end{figure}

Also, for zero field $h$ at criticality one has (see Fig. \ref{Dt_vs_h}):
\beq
D_t(\delta\rho=0,h) \sim h^{\beta/\sigma}
\eeq
and a linear regression analysis gives $\beta/\sigma=0.302(4)$, therefore $\sigma=2.11(4)$.
Both two exponents are compatible with the previously determined ones for the DLG.



\end{document}